\documentclass[prd,aps,twocolumn,tightenlines,notitlepage,nofootinbib,preprintnumbers,letterpaper,superscriptaddress]{revtex4-2}  
\usepackage{epsfig}
\usepackage{amsfonts}
\usepackage{amsmath}
\usepackage{graphicx,mathtools,tabu}
\usepackage{url}
\usepackage{xcolor}
\usepackage{amssymb,slashed,color,graphicx,bm}
\usepackage[T1]{fontenc}
\usepackage{relsize}
\usepackage{hyperref}

\hypersetup{colorlinks,citecolor= nicegreen,linkcolor= nicered}
\definecolor{nicered}{rgb}{0.7,0.1,0.1}
\definecolor{nicegreen}{rgb}{0.1,0.5,0.1}

\def\be   {\begin{equation}}  
 \def\ee   {\end{equation}}
 \def\ba   {\begin{array}}     
  \def\ea   {\end{array}}
 \def\bea  {\begin{eqnarray}}  
  \def\eea  {\end{eqnarray}}
 \def\bean {\begin{eqnarray*}}  
 \def\eean {\end{eqnarray*}}

  \def\be {\beta}

\def \R  {R_{\rm CCSN}}

\def\to {\rightarrow}

\newcommand{\sbdm}{SnBDM }

\newcommand{\km}{{\rm\ km}}
\newcommand{\cmsq}{{\rm\,cm^2}}
\newcommand{\keV}{{\rm\ keV}}
\newcommand{\MeV}{{\rm\ MeV}}

\newcommand{\phichi}{\phi_\chi}

\newcommand{\Phinu}{\Phi_\nu(E_\nu)}

\newcommand{\mchi}{m_\chi}

\newcommand{\sce}{\sigma_{\chi e}}
\newcommand{\nel}{n_\mathrm{e}}
\newcommand{\mel}{m_\mathrm{e}}

\newcommand{\dd}{\text{d}}

\begin{document}

\title{Boosted dark matter from diffuse supernova neutrinos}

\author{Anirban Das}
\email{anirband@slac.stanford.edu}
\affiliation{SLAC National Accelerator Laboratory, Stanford University, 2575 Sand Hill Road, Menlo Park, CA 94025, USA}
\author{Manibrata Sen}
\email{manibrata@berkeley.edu}
\affiliation{Department of Physics, University of California Berkeley, Berkeley, California 94720, USA}
\affiliation{Northwestern University, Department of Physics \& Astronomy, 2145 Sheridan Road, Evanston, IL 60208, USA}
\preprint{SLAC-PUB-17591, N3AS-21-005}
\date{June 11, 2021}

\begin{abstract}
The XENON collaboration recently reported an excess of electron recoil events in the low energy region with a significance of around $3.3\sigma$. An explanation of this excess in terms of thermal dark matter seems challenging. We propose a scenario where dark matter in the Milky Way halo gets boosted as a result of scattering with the diffuse supernova neutrino background. This interaction can accelerate the dark-matter to semi-relativistic velocities, and this flux, in turn, can scatter with the electrons in the detector, thereby providing a much better fit to the data. We identify regions in the parameter space of dark-matter mass and interaction cross-section which satisfy the excess. Furthermore, considering the data only hypothesis, we also impose bounds on the dark-matter scattering cross-section, which are competitive with bounds from other experiments. 
\end{abstract}

\maketitle
%
%
%
%
\section{Introduction}
Dark Matter (DM) constitutes a whopping one-fourth of the energy budget of the Universe, yet we remain in the dark about its origin and nature. The vanilla DM candidate is expected to have little or no interactions with ordinary matter beyond gravity, and hence poses a challenge to detect. Battling the odds, the scientific community has made vast progress in the past few decades, both in terms of theory, as well as detection techniques~\cite{Bertone:2004pz}. While no conclusive evidence has emerged, in their recent Science Run 1 (SR1) data, the XENON1T experiment has seen an excess of electron-recoil events in the low-energy data $(2-3\,{\rm keV})$~\cite{Aprile_2020}. More specifically, in the energy range $1-7\keV$, 285 events have been observed compared to an $232\pm 15$ events expected from the background-only model. This $3.3\sigma$ excess in the electron-recoil events has stirred up quite an excitement.

The XENON collaboration has offered three possible beyond-the-Standard-Model (BSM) explanations of this excess-- (i) solar axions, (ii) anomalous neutrino magnetic moments, and (iii) bosonic dark matter~\cite{Aprile_2020}. However, the first two solutions run into trouble with astrophysical cooling bounds~\cite{2019arXiv191010568A}, while the third solution produces a mono-energetic peak instead of an excess in a few bins. Other possible origin of the excess is attributed to unresolved $\beta$-decays of tritium~\cite{Aprile_2020}. A host of other papers followed suit with various intriguing BSM explanations~\cite{Smirnov:2020zwf,Bell:2020bes,Davoudiasl:2020ypv,Boehm:2020ltd,Arcadi:2020zni,Borah:2020jzi,Chala:2020pbn,Lindner:2020kko,DiLuzio:2020jjp,Bally:2020yid,Gao:2020wer,Bramante:2020zos,Cacciapaglia:2020kbf,Miranda:2020kwy,Davoudiasl:2020ypv,Karozas:2020pun,Zu:2020bsx,Farzan:2020llg,Khan:2020vaf,Dutta:2021nsy,McKeen:2020vpf,Jho:2021rmn}. 
All these works emphasized that an explanation in terms of the usual thermal DM seems difficult, owing to its non-relativistic nature. 

So, is a DM interpretation of the signal completely ruled out? Recent works have demonstrated that the XENON1T excess can be vindicated by a component of dark matter that gets boosted towards the earth~\cite{Kannike:2020agf,Fornal_2020,Alhazmi:2020fju}. The idea of boosted dark matter (BDM) is not new; a number of mechanisms have been put forward to render a non-thermal component to the DM velocity by scattering with cosmic rays, or annihilation of heavier dark candidates~\cite{DEramo:2010keq,Agashe:2014yua,Berger:2014sqa,Ema:2018bih,Bringmann:2018cvk,Cappiello:2018hsu,Yin:2018yjn}. Such a BDM can lead to more energetic signatures in current as well as upcoming direct detection and neutrino experiments~\cite{Bringmann:2018cvk,McKeen:2018pbb,Dent:2019krz,Ema:2020ulo}.

DM boosted by neutrinos is, on the other hand, a relatively new concept. Neutrinos interact very weakly with other SM particles, and are most elusive among all particles. This naturally makes one wonder if neutrinos have any secret interactions with DM. If true, then such interaction could alter the energy spectrum of the DM particles.  Implications of such processes have been studied in a few cosmological and astrophysical contexts~\cite{Campo:2017nwh,Ghosh:2019tab, Zhang:2020nis, Jho:2021rmn}. It is also interesting to study scenarios where DM interacts only with the leptons in SM\,\cite{Fox_2009,Bernabei_2008,Chang_2014}.

Our universe is bathed in a sea of MeV-neutrinos, emerging from massive stars going supernova, right from the epoch of first star formation. The diffuse supernova neutrino background (DSNB) is an isotropic flux of neutrinos and antineutrinos of all flavors, emitted from all core-collapse supernovae (CCSNe) in our Universe~\cite{Beacom:2010kk,Lunardini:2010ab}. The DSNB provides a perfect astrophysical laboratory to test fundamental neutrino physics~\cite{deGouvea:2020eqq}. 

In this paper, we consider the novel idea that the DM in the Milky Way (MW) halo experiences scattering with the DSNB, and gets boosted to semi-relativistic velocities. Without alluding to any specific models, we envision a scenario where DM couples to neutrinos and electrons with equal strength. Such supernova-boosted-dark-matter (\sbdm) can leave interesting signatures in low-energy recoil experiments, and even account for the excess observed in XENON1T. Furthermore, considering the XENON1T low energy data, we impose constraints on the DM-electron cross-section. Implications of such interactions of lepotophilic DM for underground detectors are also discussed.
\section{DSNB spectra}
Estimating the DSNB flux at Earth requires a precise knowledge of the rate of CCSN, as well as the flavor-dependent neutrino spectra emitted from a SN. The CCSN rate (denoted by $ \R(z)$)  is proportional to the rate of star formation (SFR), and has been well estimated by astronomical surveys~\cite{Hopkins:2006bw}. We follow \cite{deGouvea:2020eqq} (and references therein) to estimate the CCSN rate. The neutrino emission from a SN follows an approximate thermal distribution, and can be parameterized by non-degenerate Fermi-Dirac distribution~\cite{Beacom:2010kk},
\begin{equation}\label{eq:Nuspec}
F_\nu(E)=\frac{E_\nu^{\rm tot}}{6}\,\frac{120}{7\pi^4}\,\frac{E_\nu^2}{T_\nu^4}\,\frac{1}{e^{E_\nu/T_\nu}+1}\,,
\end{equation}
where $T_\nu$ is the flavor dependent neutrino temperature, and $E_\nu^{\rm tot}$ is the total energy emitted in the form of neutrinos. As neutrino emission is dominated by the cooling phase, lasting for at least $10\,$s post-bounce, it is reasonable to expect that the total emitted energy is equipartitioned among all the flavors. 

The DSNB flux can be calculated as~\cite{Horiuchi:2008jz,Beacom:2010kk}
\begin{equation}\label{eq:DSNB}
\Phi_{\nu}(E)=\int_0^{z_{\rm max}}\frac{dz}{H(z)}\R(z) F_\nu(E)\,,
\end{equation}
where $H(z) = H_0\, \sqrt{\Omega_m (1+z)^3+\Omega_\Lambda}$ is the Hubble function with $H_0=67.36\,{\rm km\,s}^{-1}\,{\rm Mpc}^{-1}$, and $\Omega_m$ and $\Omega_\Lambda$ give the matter and vacuum contribution to the energy density ~\cite{Aghanim:2018eyx}. The energy observed at the earth is redshifted such that $E=E'(1+z)$, while the maximum redshift of star-formation is taken to be $z_{\rm max}\sim 5$.

\begin{figure}[!t]
    \centering
    \includegraphics[width=0.47\textwidth]{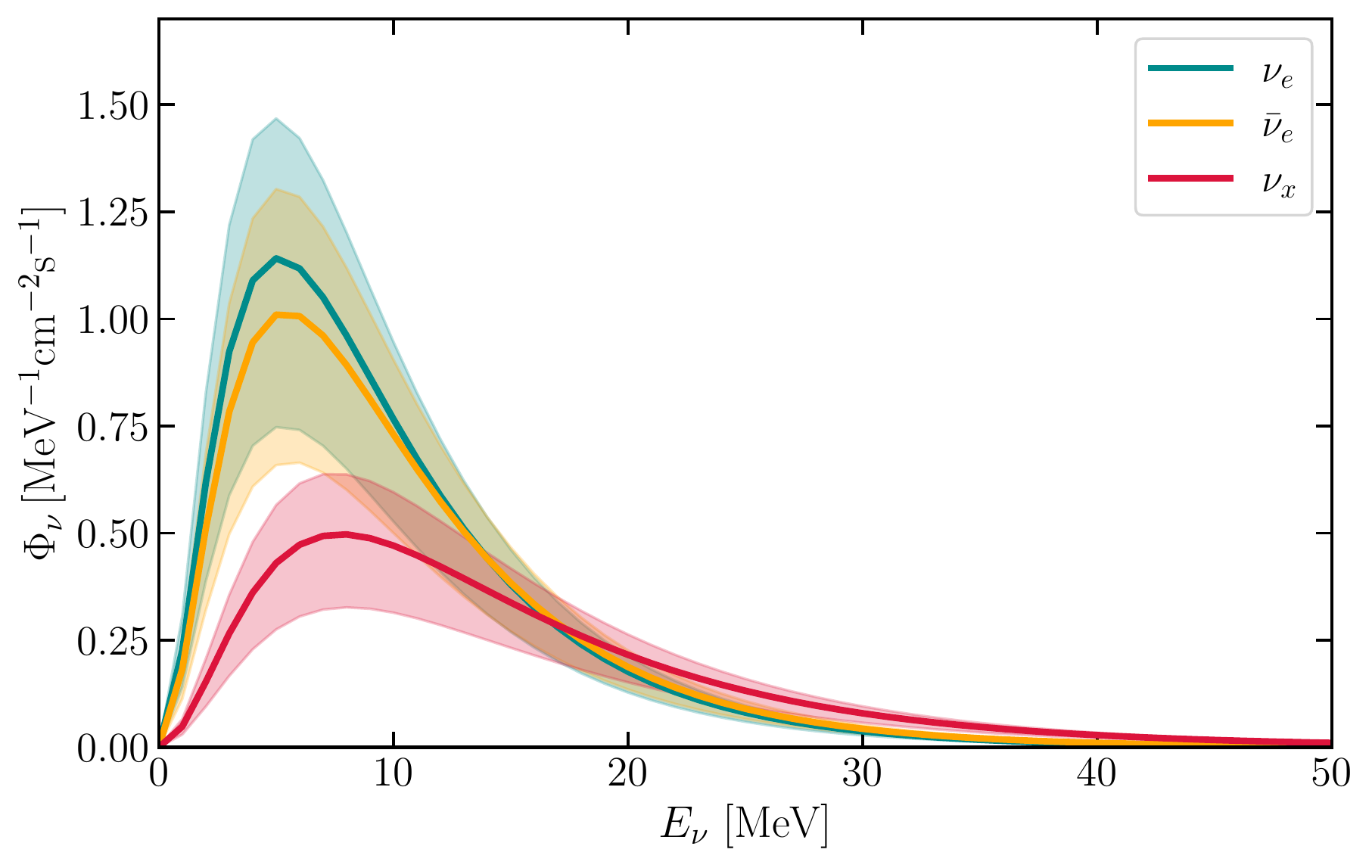}
    \caption{DSNB flux for each neutrino species. The temperatures are taken to be $T_{\nu_e}=6\MeV$,$T_{\overline{\nu}_e}=7\MeV$ and $T_{\nu_x}=10\MeV$, consistent with the bounds imposed by Super-Kamiokande~\cite{Zhang:2013tua}. The shaded regions represent the uncertainties in the DSNB spectra. The range of values for $\R(z)$ is taken from \cite{Horiuchi:2008jz}.}
    \label{fig:DSNB}
\end{figure}

We show the DSNB flux for all three flavors in Fig.\,\ref{fig:DSNB}, with the uncertainties stemming from the SFR. Due to charged current interactions, $\nu_e,\overline{\nu}_e$ have a lower temperature than $\nu_{x=\mu,\tau}$, leading to a larger sensitivity for the electron flavor neutrinos. Note that we only assume adiabatic Mikheyev-Smirnov-Wolfenstein resonant oscillations in effect inside the SN~\cite{PhysRevD.17.2369,Mikheev:1986gs}, and neglect the more uncertain collective oscillations~\cite{Duan:2006an,Hannestad:2006nj} and shock-related effects~\cite{Dasgupta:2005wn}. In the energy range we are interested in, the neutrino fluxes are fairly similar, and hence outcomes due to collective oscillations would not change our results by more than $\mathcal{O}(10\%)$~\cite{Lunardini:2010ab}. Given the uncertainty in the SFR is of $\mathcal{O}(40\%)$~\cite{Horiuchi:2008jz}, neglecting collective oscillations is a fair assumption. We operate in the normal mass ordering, where all the $\nu_e$ exit the CCSN as a $\nu_3$, while the other flavors exit as incoherent superpositions of the remaining mass states.
\section{Boosted Dark Matter}
Without referring to any particular model, we assume that the DM particle $\chi$, of mass $\mchi$, couples to electrons and neutrinos with a cross-section $\sce$. This type of scenario can arise in several leptophilic neutrino-portal models of particle DM\,\cite{Bernabei_2008,falkowski2009dark,Fox_2009,Chang_2014,Lindner_2010,macias2015effective,Blennow_2019,FileviezPerez:2019cyn}.
For concreteness, we assume that the mediator is heavier than typical energy transfer of a few $\keV$ to the electrons, but lighter than $\mchi$. This allows for large values of the DM-electron cross-section for reasonably small couplings. 

The mechanism of boosting DM particles with neutrinos is similar to that of cosmic-ray boosted DM, which has been studied extensively in the literature~\cite{Bringmann:2018cvk,Cappiello:2018hsu}. We follow their work to compute the \sbdm flux.
The DSNB neutrinos have energies of order $E_\nu\sim \mathcal{O}(10\MeV)$. The DM particles in the halo can be assumed to be at rest relative to the neutrinos. This implies that the initial virial velocity of DM is not relevant since collisions with neutrinos will impart a much higher velocity to it. We shall also neglect the neutrino mass throughout this work. With these assumptions, the energy $T_\chi$ transferred to $\chi$ by a neutrino in a single scattering is given by
\begin{align}
	T_\chi = T_\chi^\mathrm{max} \left(\frac{1-\cos\theta_\mathrm{CM}}{2}\right)\,,\\
	T_\chi^\mathrm{max} = \frac{E_\nu^2}{E_\nu+\mchi/2}\,.\nonumber
\end{align}
Here $\theta_\mathrm{CM}$ is the scattering angle in the center-of-mass frame, and $T_\chi^\mathrm{max}$ is the maximum energy that can be transferred by a neutrino of energy $E_\nu$.
\begin{figure}[t]
    \centering
    \includegraphics[width=0.47\textwidth]{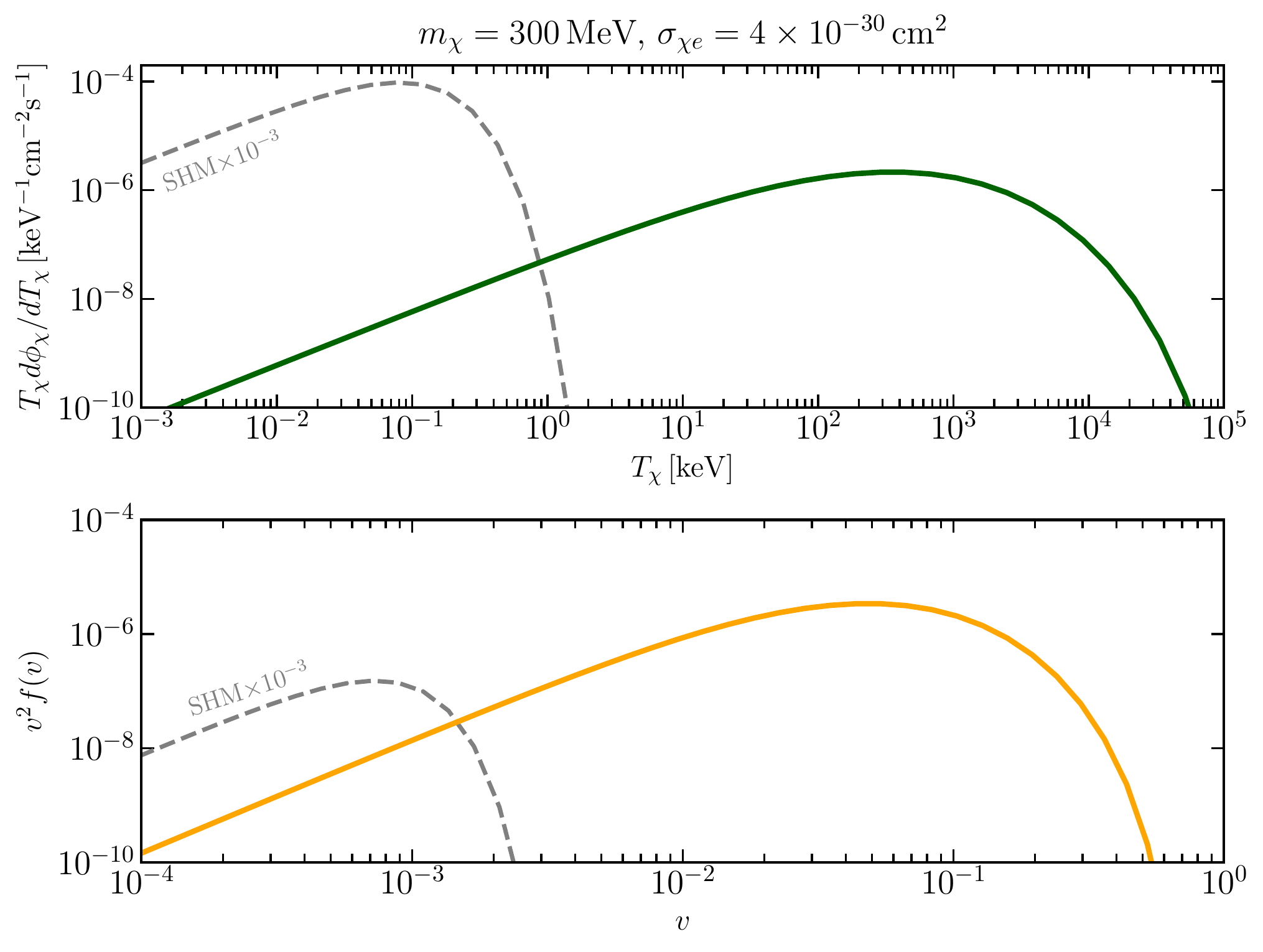}
    \caption{The \sbdm flux and velocity distribution for $\mchi=300\MeV$ and $\sce=4\times 10^{-30}\cmsq$. The rescaled Standard Halo Model (SHM) velocity distribution is shown in dashed gray line.}
    \label{fig:bdm_flux}
\end{figure}

The DM density $\rho_\chi(r)$ in the MW halo is assumed to be given by a Navarro-Frenk-White (NFW) profile where $r$ is the radial distance from the Galactic center (GC)\,\cite{Navarro_1996}. Looking from the earth, the DM density is a function of both line-of-sight (los) distance $l$ and its direction $\Omega$. Therefore, the total upscattered flux of the DM integrated over the whole MW halo is given by,
\begin{equation}\label{eq:bdm_flux1}
	\frac{d\phichi}{dE_\nu} = \int \frac{d\Omega}{4\pi} \int dl \frac{\rho_\chi}{\mchi} \Phinu \sce = D_\mathrm{halo} \Phi_\nu(E_\nu) \frac{\sce}{\mchi}\,.
\end{equation}
Here, $\Phinu$ is the DSNB flux computed in the previous section. In the last step, we have factored the whole quantity into three mutually independent parts: $D_\mathrm{halo}$ encodes the los and the angular integrals over the DM density $\rho_\chi(r)$ and depends only on the halo profile, $\Phinu$ is the DSNB flux, and $\sce,\,\mchi$ represent the new BSM parameters of the theory. We find $D_\mathrm{halo}=2.02\times 10^{25}\MeV\,\mathrm{cm^{-2}}$.
Using Eq.(\ref{eq:bdm_flux1}), the boosted DM flux can be written as
\begin{equation}\label{eq:bdm_flux2}
	\frac{d\phichi}{dT_\chi} = \int dE_\nu \frac{d\phichi}{dE_\nu} \frac{1}{T_\chi^\mathrm{max}(E_\nu)} \Theta\left[T_\chi^\mathrm{max}(E_\nu)-T_\chi\right]\,.
\end{equation}
Here, the Heaviside step function $\Theta\left[T_\chi^\mathrm{max}(E_\nu)-T_\chi\right]$ ensures that for a given neutrino energy, there is a maximum energy upto which the DM can be boosted. For a heavy-mediator interaction, the differential scattering cross-section yields a flat distribution. Fig.\,\ref{fig:bdm_flux} shows an example flux of the \sbdm for $\mchi=300\,\MeV$ and $\sce=4\times 10^{-30}\cmsq$. It is quite evident that the DM flux gets boosted by orders of magnitude, compared to that in the Standard Halo Model (SHM).

The boosted DM, after traveling through the MW, scatters with the electrons in the detector at the earth. The formalism for DM-$e$ scattering is similar to scattering with neutrinos with the following replacements: $\nu\to\chi$, and  $\chi\to e$. As mentioned earlier, we assume the scattering cross-section to be the same as in case of DM-$\nu$ scattering. Phenomenological implications of this will be discussed later.

Following Eq.(\ref{eq:bdm_flux2}), we can write the electron recoil spectrum
\begin{align}\label{eq:e_recoil}
    \dfrac{d\Gamma}{dT_e} = \dfrac{Z_\mathrm{Xe}}{m_\mathrm{Xe}} \int dT_\chi \dfrac{d\phichi}{dT_\chi} \dfrac{1}{T^\mathrm{max}_e(T_\chi)}\sce\,.
\end{align}
Here, $Z_\mathrm{Xe}=40$ is the \emph{effective atomic number} of xenon and takes care of the fact that all electrons in a xenon atom are not available for ionization, $m_\mathrm{Xe}$ is the mass of a xenon atom, and $d\phichi/dT_\chi$ is given by Eq.(\ref{eq:bdm_flux2})\,\cite{Aprile_2020}. We further convolute this electron recoil spectrum with a Gaussian detector response function with a width
\begin{equation}
    \sigma(E) = a\sqrt{E} + bE\,,
\end{equation}
with $a=0.31\sqrt{\text{keV}}, b=0.0037$\,\cite{Aprile_2020}. The XENON1T SR1 data and the background model spectrum $B_0$ are taken from Ref.\,\cite{Aprile_2020}.

\begin{figure}[!t]
    \centering
    \includegraphics[width=0.47\textwidth]{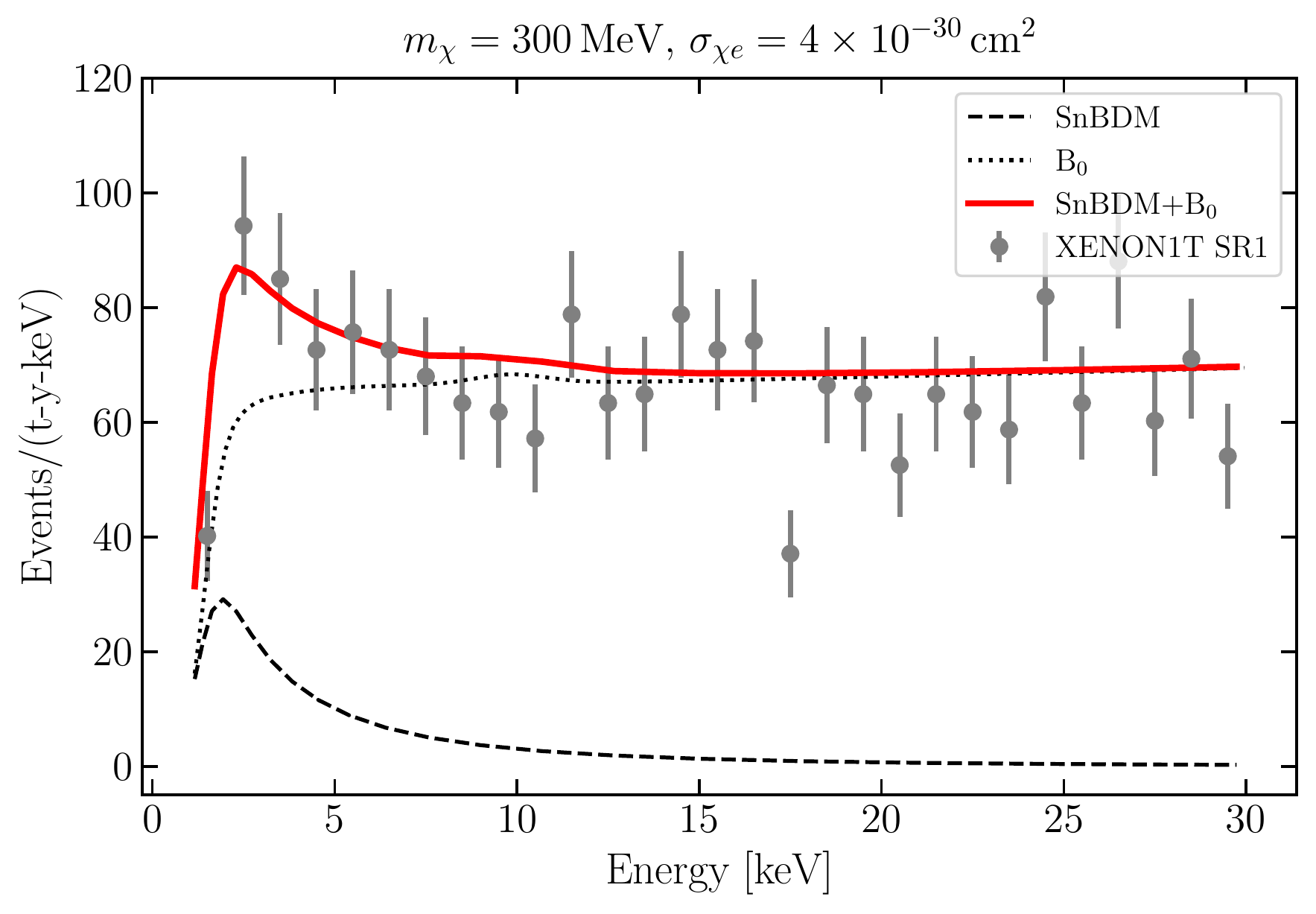}
    \caption{The electron recoil spectrum (red, solid) in XENON1T for $\mchi=300\MeV$ and $\sce=4\times 10^{-30}\cmsq$. The black dashed curve represents the contribution from \sbdm only. The black dotted curve represents the background model $B_0$\,\cite{Aprile_2020}.}
    \label{fig:recoil}
\end{figure}

Using these details, the corresponding recoil spectrum is shown in Fig.\,\ref{fig:recoil} for a sample value of $\mchi=300\,\MeV$ and $\sce=4\times 10^{-30}\cmsq$. Clearly, the recoil spectra predicted by the \sbdm model+$B_0$ (red line) presents a much better fit to the XENON1T data than the $B_0$-only hypothesis. In the following section, we perform a detailed scan of the parameter space of \sbdm to identify regions which can explain the excess.

\section{Methodology \& Results}
We utilize the SR1 data in two ways. Given the $\sim 3\sigma$ excess in electron recoil events in the $2-5\keV$ region, we try to explain the excess in terms of the SnBDM, and identify regions in the $\mchi-\sce$ parameter space that yields a better fit to the data than the $B_0$. Furthermore, using only the XENON data, we also do a general exclusion analysis. These are explained in detail below\footnote{The codes used for the calculations in this paper and the jupyter notebooks used to generate the plots can be found at this \href{https://github.com/anirbandas89/SnBDM}{https url}.}.

\begin{figure}[t]
    \centering
    \includegraphics[width=0.47\textwidth]{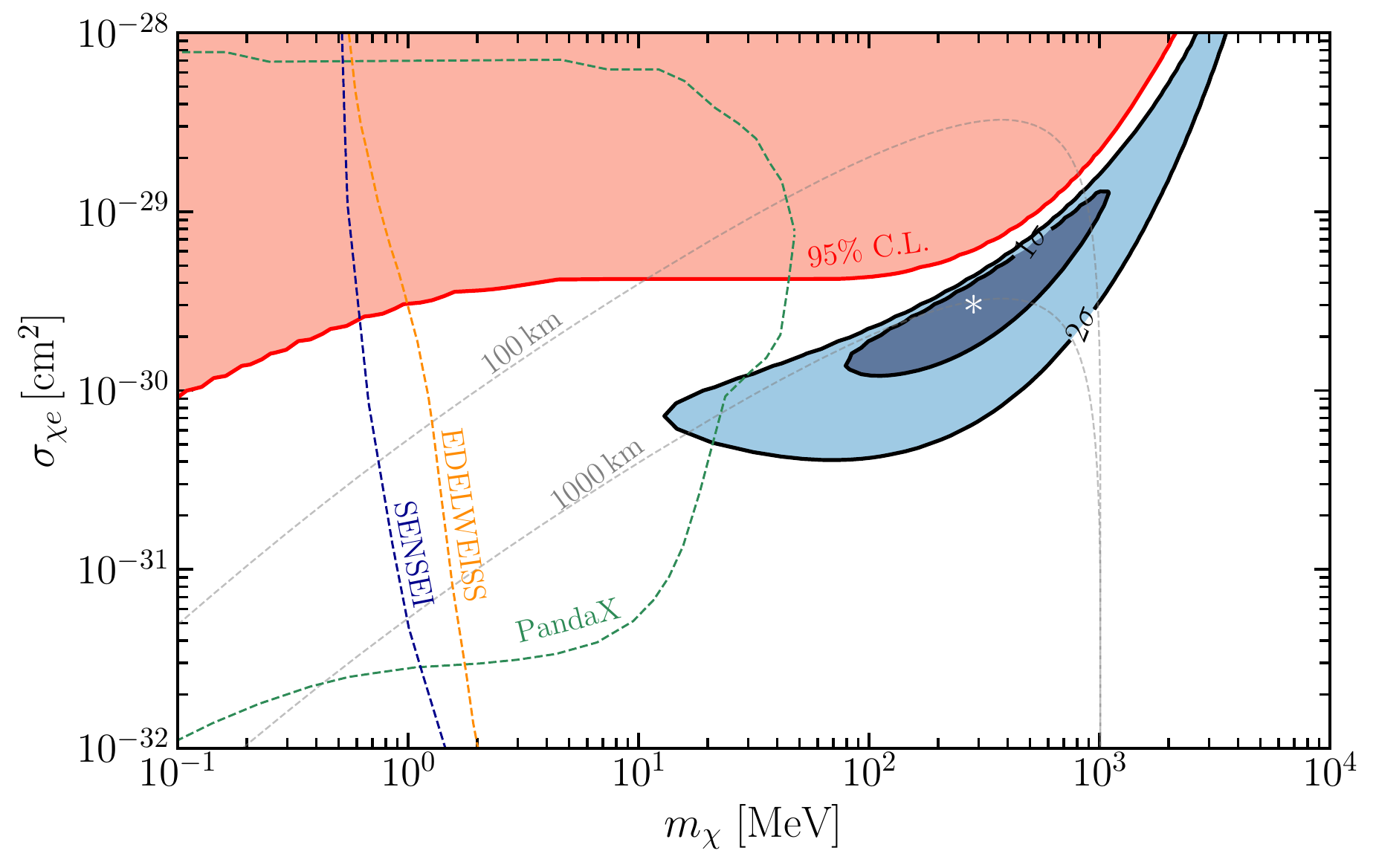}
    \caption{Contours in the $\mchi-\sce$ plane, depicting the regions which satisfy the XENON1T excess, as well as exclusion contours. The 1 and $2\sigma$ confidence intervals satisfying the excess are shown in blue contours, with the best-fit point marked with a white asterisk. The region above the red line is excluded at $95\%$ confidence level. The dashed gray lines denote the cross sections for which a DM particle travels a distance of $100\km$ and $1000\km$, respectively, before its energy falls below the threshold of XENON1T experiment. We also show the exclusion limits from SENSEI (dashed blue), EDELWEISS (dashed orange), and PandaX (dashed green) experiments.}
    \label{fig:contour}
\end{figure}

We define the following test statistic
\begin{equation}
    \chi^2 = \sum_i\dfrac{\left(D_i - M_i(\mchi,\sce)\right)^2}{\sigma_i^2}\,,
\end{equation}
where, $D_i$ is the XENON data, $M_i$ denotes the total number of events predicted by the the \sbdm model and the background, and $\sigma_i^2\equiv M_i+\sigma_{Di}^2$ is the combined uncertainty in  model and data, with $i$ running over all energy bins in the electron recoil spectrum. For simplicity, we neglect the uncertainty arising from SFR in this as well as subsequent analyses. However, we have explicitly checked that the $40\%$ uncertainty in DSNB changes the best-fit $\chi^2$ by about $0.4$. The resulting $\chi^2$ plots corresponding to $1\sigma$ and $2\sigma$ confidence limits are shown in blue contours in Fig.\,\ref{fig:contour}. The best-fit point with $\mchi=286\MeV$ and $\sce=3\times 10^{-30}\cmsq$, marked with a white asterisk, corresponds to a  $\Delta\chi^2=-5.1$ relative to the background model $B_0$. The contours show a positive correlation between $\mchi$ and $\sce$. This can be understood as follows: heavier DM mass decreases the probability of electron recoil, so larger cross section is needed to reproduce the observed spectrum.

We also compute the values of the cross section $\sce$ that can be excluded given the observed electron recoil. To this end, we take the following conservative approach to construct a test statistic $\chi_>^2$\,\cite{Ackermann:2015tah},
\begin{equation}
    \chi_>^2 = \sum_i \dfrac{\left(D_i - M_i(\sce)\right)^2}{\sigma_i^2}\,.
\end{equation}
Here, we sum over only those energy bins where the model $M_i$ alone predicts more events than what is observed. This is a conservative method and stays completely agnostic about any background model. For each value of $\mchi$, we scan over $\sce$ to compute $\chi_>^2$. This allows us to compute an exclusion contour at $95\%$ confidence limit when $\chi_>^2 \ge 4$. This limit is shown by a red contour in Fig.\,\ref{fig:contour}. All cross sections above this line are excluded at $2\sigma$ significance. By construction, this does not have an overlap with the best-fit contours at $2\sigma$.

The detector of the XENON experiment sits in an underground facility about 1.4 km below Earth's surface. Therefore, if the DM-electron cross section is too large, the DM particles could lose most of its energy and be stopped in the earth before it reaches the detector. Following Ref.\,\cite{PhysRevD.41.3594}, we perform a simple energy loss analysis to estimate this. If a DM particle has kinetic energy $T_\chi$ and it scatters with the electrons, then its energy loss rate is given by
\begin{equation}\label{eq:overburden1}
    \dfrac{dT_\chi}{dx} = -\nel \int_0^{E_R^\mathrm{max}} \dd E_R E_R \dfrac{d\sce}{dE_R}\,.
\end{equation}
Here, $\nel$ is the electron number density in Earth, $E_R$ is the electron recoil energy in a single collision. This relation can be simplified by performing the $E_R$ integration in the heavy mediator limit, $d\sce/dE_R=\sce/E_R^\mathrm{max}$. This leads to the following expression for the final energy of the DM particle after traveling a distance $x$,
\begin{equation}\label{eq:overburden2}
    T_\chi(x) = T_\chi^\text{ini}\exp\left[-\dfrac{2\nel\sce \mel x}{\mchi}\right]\,.
\end{equation}
The XENON experiment has a low energy threshold of $\simeq 2\keV$ for electron recoil\,\cite{Aprile:2019dme}. Now Eq.(\ref{eq:overburden2}) can be inverted to find out the threshold cross section $\sce^\text{th}$ for which a DM particle can travel a distance $x$ before its energy falls below the XENON detector threshold.
Fig.\,\ref{fig:contour} shows two gray dashed lines for $\sce^\text{th}$ corresponding to distances $x=100$\,km, $1000$\, km. A more detailed analysis would involve integrating $\sce^\text{th}$ over  all energies weighted by the flux spectrum. However, we assumed $T_\chi^\text{ini} \sim 1\MeV$ for the sake of simplicity as $\sce^\text{th}$ has only logarithmic dependence on $T_\chi^\text{ini}$. We see that the traveling distance of the DM particles inside the earth is far greater than the depth of the XENON experiment for $\sce \sim 10^{-30}-10^{-29}\cmsq$. Nevertheless, this will imprint a directionality in the DM flux. The flux coming through the bulk of the earth will be more attenuated compared to the flux coming from above. In Fig.\,\ref{fig:contour}, we also show the exclusion limits from SENSEI (dashed blue), EDELWEISS (dashed orange), and PandaX (dashed green) experiments\,\cite{Barak:2020fql, Arnaud:2020svb, Jho:2021rmn}.
\section{Daily Modulation of Dark Matter Signal}
As the Earth rotates about its axis, the boosted DM flux coming from a specific direction in the sky has to travel a varying distance through the Earth. This will cause a daily modulation in the DM scattering event rate in a detector. Such daily modulation in DM signal has been calculated before, for example, in Ref.\cite{Kouvaris:2014lpa}. In this section, we shall provide a rough estimate of such a daily modulation in the \sbdm{} signal.

Let us consider a reference frame where the $z$-axis is along the north-south pole of the Earth. The DM wind is assumed to be coming from a direction with a latitude $\approx 42^\circ$. Therefore, $\theta \approx 48^\circ$ is the corresponding polar angle. The velocity can be written as
\begin{equation}
    \Vec{v}_\chi = v_\chi (\hat{x} \sin\theta \cos\phi + \hat{y} \sin\theta \sin\phi + \hat{z} \cos\theta)\,.
\end{equation}
We assume that the DM particles retain this directionality even after scattering with the DSNB because the interaction mediator is heavier than typical momentum exchange, resulting in isotropic scattering. We shall ignore the velocity of the Earth compared to that of the DM particles, which are boosted to $\sim \mathcal{O}(1\MeV)$ energy after scattering. If $\hat{n}$ denotes the direction of the XENON detector, then
\begin{equation}
    \hat{n} = \hat{x} \cos{\theta_d} \cos{\omega t} + \hat{y} \cos{\theta_d} \sin{\omega t} + \hat{z} \sin{\theta_d}\,,
\end{equation}
where $\theta_d$ is the latitude of the detector and $\omega$ is the angular velocity of the Earth. The angle $\psi$ between the detector and the direction of DM wind is given as
\begin{equation}
    \cos\psi = \hat{v}_\chi\cdot\hat{n} 
\end{equation}
Using the angle $\psi$, we can find the distance travelled by the DM particle through the Earth,
\begin{equation}
    L_\chi = (R_E - L_d)\cos\psi + \sqrt{(R_E - L_d)^2\cos^2\psi - L_d^2 + 2R_EL_d}\,,
\end{equation}
where $R_E$ is the radius of the Earth and $L_d=1.4\km$ is the depth of the location of the detector.

\begin{figure}[!t]
    \centering
    \includegraphics[width=0.48\textwidth]{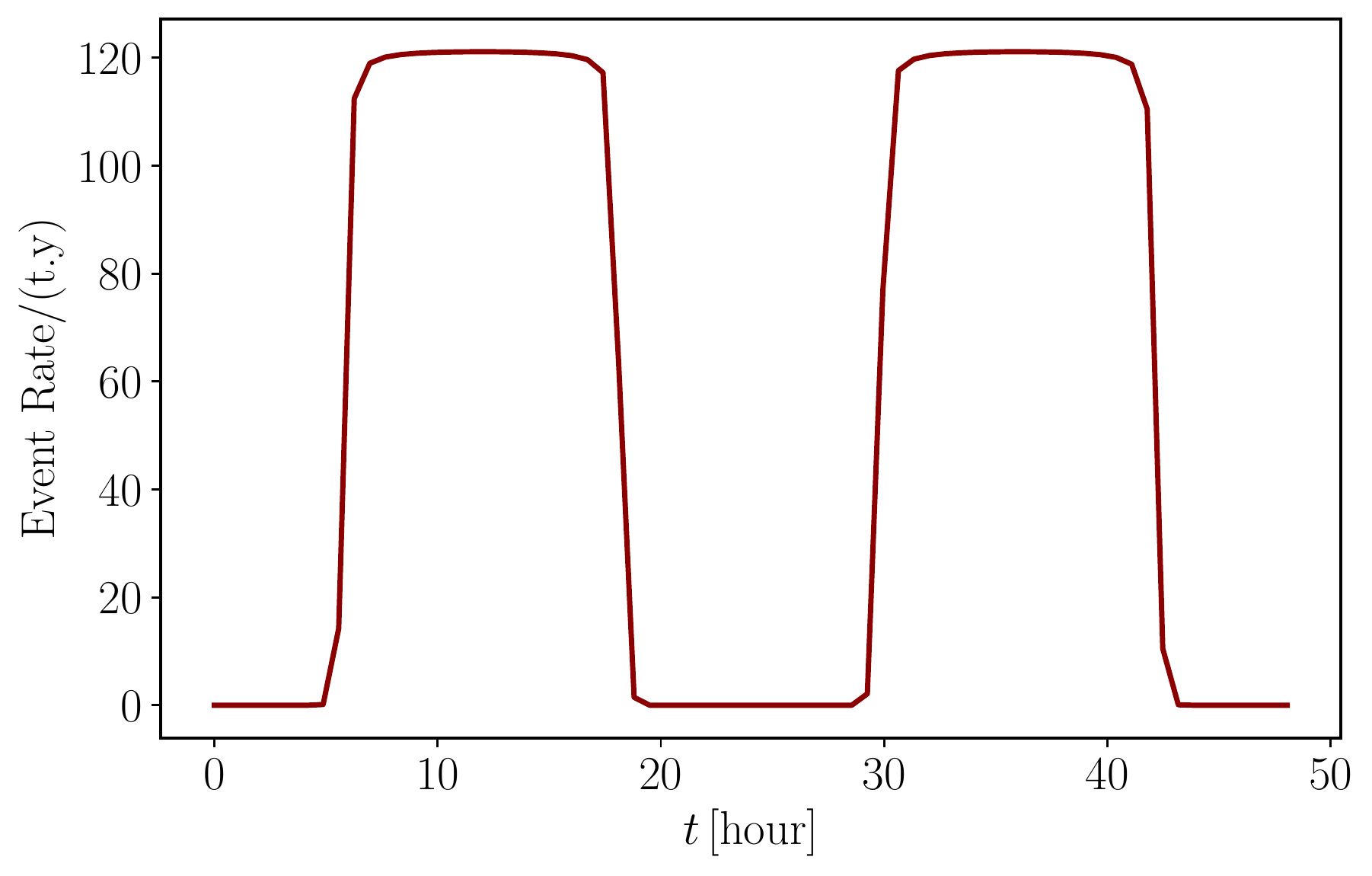}\\\includegraphics[width=0.48\textwidth]{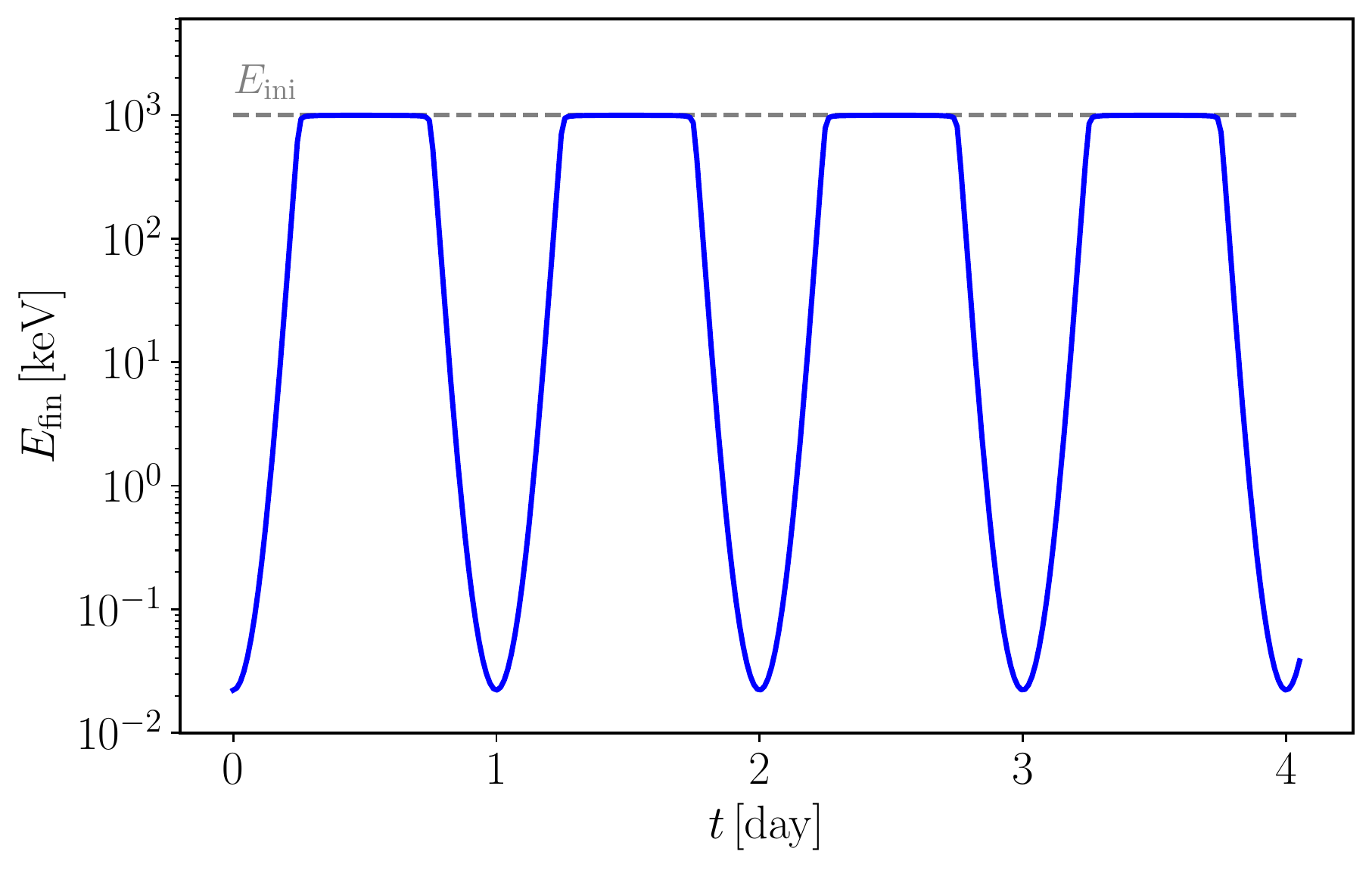}
    \caption{(Left) The total electron-recoil rate variation with time. (Right) The time variation of the final DM energy at the XENON detector assuming an initial energy at the surface of the Earth $E_\mathrm{ini}=1\MeV$.}
    \label{fig:vars}
\end{figure}

DM particles lose energy due to scattering with the electrons in the medium while travelling inside the Earth. When the interaction mediator is heavier than the typical momentum exchanged (heavy mediator limit), the final energy of the DM particle after travelling a distance $L_\chi$ is given by
\begin{equation}
    T_\chi = T_\chi^\text{ini} \exp\left[-\frac{2n_e\sce \mel L_\chi}{\mchi}\right]\,.
\end{equation}
A daily modulation in $L_\chi$ will impart a modulation in DM kinetic energy $T_\chi$, which will in turn imply a 24-hour period variation in the event rate. In Fig.\,\ref{fig:vars}, we show this time variation in the total expected event rate and DM energy after passing through the Earth. Clearly, one can notice a significant daily modulation in the event rate, as well as the final DM energy, and this can be used to put constraints on the DM-electron coupling.

\section{Conclusion \& Outlook }
The observation of the electronic recoil excess in the XENON1T experiment has renewed the interest in boosted DM. 
In this work, we present a scenario where DM couples to neutrinos and electrons. The Universe is permeated by a constant source of  MeV-energy neutrinos originated from all past supernovae, which may boost the DM through scattering. This can be a lucrative explanation for the excess of events observed by the XENON1T experiment. We also perform a conservative analysis to constrain large DM-electron cross sections without referring to the excess as such.

DM-electron interactions also lead to a shielding effect as the DM travels through the earth. The flux is attenuated more when it travels a longer distance through the earth to reach the detector. This creates a diurnal modulation in the event rate in the detector~\cite{Fornal_2020}. The resulting modulation in the total event rate is computed. If such a signal is observed, the amplitude of the modulation could be used to determine the DM interaction strength with ordinary matter irrespective of the total number of recoil events. This will be an interesting direction for future investigation.

Furthermore, such BSM interaction with leptons could have other observable effects. For example, the scattering with DM will change the DSNB energy spectrum. The neutrinos will lose energy by upscattering the DM particles, and the DSNB spectrum as shown in Fig.\,\ref{fig:DSNB} will shift towards lower energy. We found that change in DSNB would be pronounced if the DM and the mediator masses $\lesssim\mathcal{O}(\text{few}\MeV)$. However, models with such small DM masses and large interaction cross section will inevitably run into conflict with BBN. We leave this exercise for a future analysis.

Neutrino telescopes are another class of experiments that are promising to detect \sbdm. The typical detection threshold for these detectors are above $\sim 10\MeV$ which is not suitable for detecting the vanilla galactic halo DM. However, if the DM is boosted to semi-relativistic speeds, then they can scatter with the electrons in the neutrino detectors and deposit an energy above threshold. The expected event rate has been computed in a few previous works and competitive constraints on DM-electron cross section have been obtained\,\cite{Bringmann:2018cvk}. One way to test the solution presented here would be to compute the expected number of events in the MiniBooNE or BOREXINO experiment and compare with the data\,\cite{Aguilar_Arevalo_2018,Aguilar-Arevalo:2017mqx,Bellini_2014}. The DM particles will also undergo collisions with the electrons in the sun, and  get scattered isotropically in all directions. As such, the sun will act as a reflector of the DM particles. This could add extra directionality in the observed DM flux at Earth.

DM-electron scattering before and during recombination would affect the cosmic microwave background (CMB) anisotropy spectrum. Previous works have constrained DM-baryon cross section using the CMB data\,\cite{Gluscevic_2018,Xu_2018}. The bounds can be translated to the present case by re-scaling the relevant interaction term with appropriate modifications. We have checked that this limit does not constrain our best-fit DM mass and cross-section. Finally, we note that such large DM-electron cross-section as presented in this work, and all other works on boosted DM, has been excluded by SENSEI, EDELWEISS, and CRESST-II experiments as shown in Fig.\,\ref{fig:contour}\,\cite{Barak:2020fql, Arnaud:2020svb, Abdelhameed:2019hmk}. 
The requirement of a mediator lighter than the electron brings up the possibility of its thermalization before BBN. However, we have explicitly checked that with mediator mass $m_{\rm med}\lesssim 0.01\MeV$ and electron coupling $g_e\lesssim 10^{-7}$, it is possible to reproduce the required DM-electron scattering cross section, and not violate the BBN constraints at the same time.

Finally, note that neutrino-boosted DM has been studied before in Refs.\,\cite{Zhang:2020nis, Jho:2021rmn} in the context of solar neutrinos, neutrinos from Sun-like stars, and cosmic-ray electrons. The constraints arising on DM-electron cross-section from these studies are typically stronger, due to the larger flux of neutrinos as compared to the DSNB. On the other hand, the DSNB represents an isotropic source of neutrinos present from a much older epoch $z\sim 5$. As a result, the constraints from this study are independent of the solar constraints.
\medskip
\section*{Acknowledgments}
We thank Basudeb Dasgupta and Andre de Gouvea for useful suggestions and comments on the manuscript. AD was supported by the U.S. Department of Energy under contract number DE-AC02-76SF00515. MS acknowledges support from the National Science Foundation, Grant PHY-1630782, and to the Heising-Simons Foundation, Grant 2017-228.

\bibliography{DM_nu}
\bibliographystyle{apsrev4-2}
\end{document}